\documentclass[12pt]{amsart} 
\usepackage{amssymb} 
\usepackage[breaklinks]{hyperref} 
\usepackage{accents} 
\usepackage{natbib} 
\usepackage{accents} 
\setcitestyle{aysep={}} 
\RequirePackage{color,graphicx} 
\newcommand{\HH}{\mathcal{H}}

\newcommand{\RR}{\mathbb{R}}
\newcommand{\CC}{\mathbb{C}}

\newcommand{\UU}{\mathcal{U}}

\newcommand\restr[2]{{
  \left.\kern-\nulldelimiterspace 
  #1 
  \vphantom{\big|} 
  \right|_{#2} 
  }}

\newcommand{\Inn}[2]{\langle #1, #2 \rangle}

\newcommand{\Norm}[1]{\left|\left\langle #1 \right\rangle \right|}

\makeatletter
\let\uppercasenonmath\@gobble
\makeatother

\newcommand{\Ss}{\mathcal{S}}
\newcommand{\Ov}{\mathcal{O}}
\newcommand{\Op}{\accentset{p}{\Ov}}

\newtheorem*{genwig}{Generalized Wigner's Principle}
\theoremstyle{definition}
\newtheorem*{defi}{Definition}

\title[Comment on Ashtekar]{Comment on Ashtekar:\\Generalization of Wigner's Principle}
\author[Bryan W. Roberts]{Bryan W. Roberts\\ University of Southern California \\
\href{http://www.usc.edu/bryanroberts}{www.usc.edu/bryanroberts} \\ \\ \today}

\begin{document}
\maketitle

\section{Introduction}

My sincere thanks to Dr. Ashtekar for his note on the roads to $T$-violation. This clarifies the situation a great deal. I have argued that when you boil down existing techniques for testing time asymmetric ($T$-violating) phenomena, you find that there are really just three principles underpinning them: Curie's principle, Kabir's principle, and Wigner's Principle \citep{roberts2013threeroads}. But is there any sense in which these principles really belong to quantum theory, or are they more general than that?

As Dr. Ashtekar illustrates with great clarity, the former two principles are significantly more general than quantum theory as we currently know it. Just how general? We don't need the dynamics to be linear, let alone unitary. We don't need a vector space or a superposition principle. We don't even need observables. Curie's principle, Dr. Ashtekar observes, follows from little more than the notion of a bijection on a set of states. And Kabir's principle can be formulated with the addition of only the bare-bones notion of an ``overlap map'' to capture some structural features of a transition probability. Both principles are true in the very minimalist formalism that Ashtekar calls \emph{general mechanics} \citep{ashtekar2013threeroads}.

Curie's Principle and Kabir's Principle are the only techniques that have led to successful tests for $T$-violation. So, these generalizations illustrate a sense in which our existing evidence for $CP$-violation and for $T$-violation is extremely robust, since these principles obtain in a variety of modifications of quantum theory.

But what about the third road to $T$-violation, Wigner's principle? My purpose in the remainder of this note is to illustrate how Wigner's Principle can be generalized as well.

\section{General Mechanics}

Following \citet{ashtekar2013threeroads}, let $\Ss$ be a set of states, and let $S:\Ss\rightarrow\Ss$ be a bijection, which we interpret as implementing dynamical evolution (like an $S$-matrix). It is helpful to think of ourselves as having two copies of that set of states denoted by different indices, $\Ss_i$ (the ``initial states'') and $\Ss_f$ (the ``final states''), and write $S:\Ss_i\rightarrow\Ss_f$. Accordingly, when it is appropriate, a single state $\sigma\in\Ss$ will be denoted with different indices $\sigma_i\in\Ss_i$ and $\sigma_f\in\Ss_f$ depending on which of the two copies it is in.

We further define an \emph{overlap map} $\Ov:\Ss\times\Ss\rightarrow\RR$ that is symmetric $\Ov(\sigma,\rho)=\Ov(\rho,\sigma)$, providing a generalization of the quantum mechanical notion of a transition probability $\Norm{\sigma,\rho}^2$. Following the index convention above, we denote this map by $\Ov_i$ when it operates on $\Ss_i\times\Ss_i$, and by $\Ov_f$ when it operates on $\Ov_f\times\Ov_f$.

\begin{figure}[bt]\begin{center}
\includegraphics[width=0.6\textwidth]{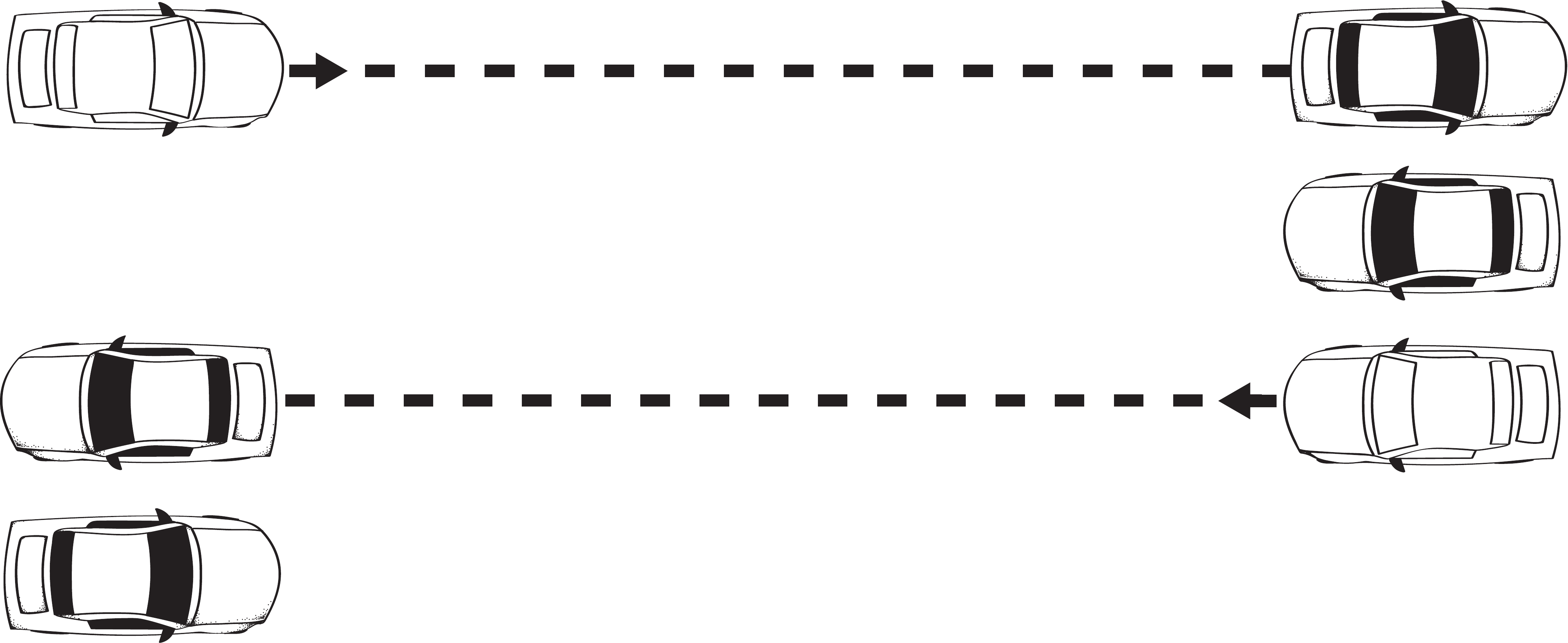}
\caption{Time reversal invariance.}\label{fig:toycar}
\end{center}\end{figure}

The \emph{time reversal operator} in general mechanics is a one-to-one mapping $T:\Ss_i\rightarrow\Ss_f$, and we interpret \emph{time reversal invariance} to mean that $T^{-1}S=S^{-1}T$. Equivalently, time reversal invariance says that $T^{-1}ST^{-1}S=I$  is the identity on $\Ss$. This captures the essential property of a time reversal invariant system that if we evolve a state, time reverse it, evolve it again, and then time reverse it again, then that is the same as if we had done nothing at all, as in Figure \ref{fig:toycar}.

\section{Degeneracy}

In quantum theory, a degenerate eigenvector $\psi$ of the Hamiltonian $H$ is one that admits an eigenvector $\phi$ that is orthogonal to it $\Inn{\psi}{\phi}=0$, but which has the same eigenvalue. Our central task in this section is to find an appropriate definition of degeneracy for general mechanics, since it is essential to the expression of Wigner's Principle.

General mechanics so far does not have enough structure to define degeneracy. We don't have linearity, so we can't talk about eigenvalues on a linear space. But can we do so using the overlap map?

To see that this is not enough, consider the special case of quantum mechanics, where the overlap map is $\Ov(\psi,\phi)=\Norm{\psi,\phi}^2$. Let $\varphi$ be a normalized energy eigenvector, and hence an eigenvector of the unitary evolution operator $\UU_t=e^{-itH}$. Then $\Norm{\varphi,\UU_t\varphi}^2=\Norm{\varphi,e^{-ith}\varphi}^2=1$ for all eigenvalues $h$. And since $\UU_t$ is unitary, $\Norm{\UU_t\psi,\UU_t\phi}^2=\Norm{\psi,\phi}$ for all $\psi,\phi$. Neither of these two calculations provides a way to distinguish eigenvectors with distinct eigenvalues. This suggests don't yet have enough information get at degeneracy. However, we can define degeneracy with the help of a little extra structure.

\subsection{Preliminary Definitions}

Let me begin by introducing two definitions, making use only of the objects that have been introduced so far.

\begin{itemize}
	\item \emph{Equivalent States} ($\mathbf{\equiv}$). Two states $\sigma,\rho\in\Ss$ are \emph{equivalent} (written $\sigma\equiv\rho$) when $\Ov(\sigma,\xi)=\Ov(\rho,\xi)$ for all $\xi\in\Ss$.
	
	In quantum mechanics, two vectors related by a complex unit represent the same state, $\psi=e^{i\theta}\phi$. This is equivalent\footnote{Proof: ($\Rightarrow$) This direction is obvious. ($\Leftarrow$) Suppose $\Norm{\psi,\xi}^2=\Norm{\phi,\xi}^2$ for all $\xi\in\HH$. Thus the complex numbers $\Inn{\psi}{\xi}$ and $\Inn{\phi}{\xi}$ have the same length, and so are related by a rotation of the complex plane. This means that for each $\xi\in\HH$, there exists some $e^{-i\theta}$ such that $\Inn{\psi}{\xi}=e^{-i\theta}\Inn{\phi}{\xi} = \Inn{e^{i\theta}\phi}{\xi}$, and hence $\Inn{\psi-e^{i\theta}\phi}{\xi}=0$. In particular, $\Inn{\psi-e^{i\theta}\phi}{\psi-e^{i\theta}\phi}=0$. Therefore, since the inner product $\Inn{\cdot}{\cdot}$ is positive definite, $\psi-e^{i\theta}\phi=0$, and so $\psi=e^{i\theta}\phi$ as claimed.} to the statement that $\Norm{\psi,\xi}^2=\Norm{\phi,\xi}^2$ for all $\xi\in\HH$. So, since $\Ov(\psi,\xi)=\Norm{\psi,\xi}^2$ in the special case of quantum theory, our definition is the natural generalization of the idea $\psi$ and $\phi$ represent the same state.
	
	\item \emph{Stationary States}. A state $\sigma\in\Ss$ is \emph{stationary} when $S\sigma\equiv\sigma$, where $S:\Ss_i\rightarrow\Ss_f$ is the dynamical evolution operator.
	
	In quantum theory, a state $\psi\in\HH$ is stationary if $\UU_t\psi=e^{i\theta}\psi$. Expressed in general mechanics with the dynamics $S$, this just says that $S\psi$ and $\psi$ are equivalent states, according to our definition above.
\end{itemize}

\subsection{Additional Structure}

We must now introduce a new structure, which I will call the \emph{pre-overlap map} $\Op$. It is a mapping $\Op:\Ss\times\Ss\rightarrow\CC$, which is taken to be compatible with both the dynamical evolution operator $S$ and the time reversal operator $T$,
\begin{align*}
	\Op(S\sigma,S\rho) & = \Op(\sigma,\rho)\\
	\Op(T\sigma,T\rho) & = \Op(\rho,\sigma),
\end{align*}
for all $\sigma,\rho\in\Ss$. This structure is thus analogous to the inner product $\Inn{\cdot}{\cdot}$ in quantum theory, but lacks much of its structure.

Our use for the pre-overlap map is to define an analogue of two states having ``the same energy eigenvalues.'' In generalized mechanics, we take this to be the statement that for stationary states $\sigma$ and $\rho$,
\begin{equation*}
	\Op(\sigma,S\sigma)=\Op(\rho,S\rho).
\end{equation*}
Again, consider the quantum analogue $\Inn{\sigma}{\UU_t\sigma}=\Inn{\rho}{\UU_t\rho}$, where $\sigma$ and $\rho$ are vectors in a Hilbert space and $\UU_t=e^{-itH}$. Since $\psi$ and $\phi$ are assumed to be stationary and normalized,
\begin{equation*}
	\Inn{\sigma}{\UU_t\sigma}=\Inn{\rho}{\UU_t\rho} \Longrightarrow e^{-ith}=e^{-ith^\prime},
\end{equation*}
where $\UU_t\sigma=e^{-ith}\sigma$ and $\UU_t\rho=e^{-ith^\prime}\rho$. This implies that $\sigma$ and $\rho$ have the same energy eigenvalues, $h=h^\prime$.

With this in mind, we can define what it means for a stationary state $\sigma$ in general mechanics to be non-degenerate. In quantum theory, non-degeneracy is the property that $\sigma$ and $\rho$ have the same energy eigenvalues only if $\sigma=e^{i\theta}\rho$. In general mechanics, this amounts to the following.

\begin{defi}
	A stationary state $\sigma\in\Ss$ is \emph{non-degenerate} if for every stationary state $\rho$, $\Op(\sigma,S\sigma)=\Op(\rho,S\rho)$ only if $\sigma\equiv\rho$. Otherwise, $\sigma$ is called \emph{degenerate}.
\end{defi}

\section{Generalization of Wigner's Principle}

With the definitions above, we now have a simple statement of Wigner's Principle.

\begin{genwig}
	Suppose there exists a stationary state $\sigma\in\Ss$ such that both of the following are true:
	\begin{enumerate}
		\item $\sigma$ is non-degenerate; and
		\item $T\sigma\not\equiv\sigma$.
	\end{enumerate}
	Then we have $T$-violation, in that $TS^{-1}\neq ST^{-1}$.
\end{genwig}
\begin{proof}
	We prove the contrapositive statement: assume $TS^{-1}=ST^{-1}$. Let $\sigma$ be stationary, $S\sigma_i=\sigma_f$, and define $\rho$ by the relations, $\rho_f:=T\sigma_i$, $\rho_i:=T^{-1}\sigma_f$. Then,
	\begin{align*}
		\Op_f(\sigma_f,S\sigma_i) & = \Op_i(T^{-1}S\sigma_i,T^{-1}\sigma_f) & \text{Compatibility of $T$}\\
								  & = \Op_i(S^{-1}T\sigma_i,T^{-1}\sigma_f) & \text{Time reversal invariance}\\
								  & = \Op_i(S^{-1}\rho_f,\rho_i)			& \text{Definition of $\rho$}\\
								  & = \Op_f(SS^{-1}\rho_f,S\rho_i)			& \text{Compatibility of $S$}\\
								  & = \Op_f(\rho_f,S\rho_i).				&
	\end{align*}
Suppose (1) is true, and $\sigma$ is non-degenerate. Then by this calculation, $\sigma\equiv\rho$. Since $\rho=T\sigma$, this implies that (2) fails. Suppose instead that (2) is true, and $T\sigma\not\equiv\sigma$. Then $\sigma\not\equiv\rho$, and (1) fails.
\end{proof}

\section{Discussion}

\citet{ashtekar2013threeroads} showed that Curie's Principle is the most general of the three roads to $T$-violation, requiring only a minimal amount of structure involving bijections on sets, while Kabir's Principle requires the addition of an overlap map. We have now seen that Wigner's Principle requires the addition of an overlap map plus a \emph{pre}-overlap map. The three roads thus require successfully more structure. Nevertheless, casting them in Ashtekar's general framework shows that they are all surprisingly robust, much more so than was understood by \citet{roberts2013threeroads}.

\bibliographystyle{kp}
\bibliography{/Users/bryanwroberts/Dropbox/ElectronicLibrary/MasterBibliography}
\end{document}